# Analysis of various climate change parameters in India using machine learning

Rutvij Wamanse
*Department of Electronics and Telecommunication.*
*Pune Institute of Computer Technology.*
Pune, India.
wamanserutvij25@gmail.com

Tushuli Patil
*Department of Electronics and Telecommunication.*
*Pune Institute of Computer Technology.*
Pune, India.
tushulipatil@gmail.com

*Abstract*—Climate change in India is one of the most alarming problems faced by our community. Due to adverse and sudden changes in climate in past few years, mankind is at threat. Various impacts of climate change include extreme heat, changing rainfall patterns, droughts, groundwater, glacier melt, sea-level rise, and many more. Machine Learning can be used to analyze and predict the graph of change using previous data and thus design a model which in the future can furthermore be used to catalyze impactful work of climate change and take steps in the direction to help India fight against the upcoming climate changes. In this paper we have analysed 17 climate change parameters about India – Forest area(square kilometer), Agricultural irrigated land (percentage of total agricultural land), Cereal yield (kilogram per hectare), Access to electricity (percentage of population), carbon dioxide intensity (kilogram per kilogram of oil equivalent energy use), carbon dioxide emissions from gaseous fuel consumption (kiloton) , carbon dioxide emissions (kiloton), carbon dioxide emissions from liquid fuel consumption (kiloton), carbon dioxide emissions (metric tons per capita), carbon dioxide emissions from solid fuel consumption (kiloton), Total greenhouse gas emissions (kiloton of carbon dioxide equivalent), hydrofluorocarbon gas emissions (thousand metric tons of carbon dioxide equivalent), Methane emissions (kiloton of carbon dioxide equivalent), Nitrous oxide emissions (thousand metric tons of carbon dioxide equivalent), Annual freshwater withdrawals, total (billion cubic meters), Total population, Urban population. We have applied linear regression, exponential regression, and polynomial regression to the parameters and evaluated the results. Using the designed model, we will predict these parameters for the years 2025,2030, 2035. These predicted values will thus help our community to prevent and take actions against the adverse and hazardous effects on mankind. We have designed and created this model which provides accurate results regarding all 17 parameters. The predicted values will therefore help India to be well equipped against climate change. This data when made available to the people of India will help create awareness among them and will help us save our country from the haphazard effects of climate change.

## I. Introduction

Earth's climate has changed throughout history. Most of these changes were slow and went unnoticed as these were majorly due to changes in Earth's orbit and the change of solar energy received by our planet. Since the mid-20th century, climate change has been proceeding at a rate that has never been seen in a millennia. Human activities have warmed the ocean, land as well as atmosphere as well as caused rapid irreversible changes in various realms of earth. India is one of the most populated countries in the world, therefore the threat faced by India and its people is fatal and if the climate changes continue at this rate soon there will be a huge crisis to face. Therefore as a citizen of India, it is our duty to contribute to curating solutions to save our country. Picciariello, A., Colenbrander, S Bazaz, A., and Roy, R. proposed various studies of disastrous flooding, deadly heatwaves, climate change, and agriculture [1]. Anil Agarwal discussed effects on agriculture, forests, people responsible for global warming and has discussed India's position at the climate negotiations [2].

The major drawback of the research done related to climate change in the past few years is that it focuses on drawing inferences based on past changes in climate and providing solutions for the present as well the conclusions obtained are not in the form of statistics. We on the other hand have used Machine Learning algorithms such as linear regression, exponential regression, and polynomial regression to provide a better understanding of the present and future using data collected. Focus is not only kept on a single parameter but 17 different parameters which will majorly affect climate changes in the future to cover all the aspects regarding climate change in India which will provide a brief understanding and therefore help us act in that direction.

The rest of the paper is as follows: Section 2 provides an overview of the overall training process which contains the description of the dataset used, the pre-processing techniques used, and the algorithms used. The evaluation of various algorithms is explained in section 3. Section 4 contains the inference and prediction values. Finally, section 5 draws the conclusion and suggests the future direction of the work.

## II. Training

### A. Dataset

The dataset contains in total of 65 climate change parameters for India for several years, but we have selected 17 parameters because other parameters were either irrelevant or had very less data points for the machine learning algorithm to work on them, for example the parameters "Land area where elevation is below 5 meters (percentage of total land area)", "Terrestrial protected areas (percentage of total land area)", "Marine protected areas (percentage of territorial waters)" has only three datapoints and all three datapoints are equal [3]. The parameter "School enrolment, primary and secondary (gross), gender parity index (GPI)" is irrelevant to climate change. The parameters "Disaster risk reduction progress score (1-5 scale; 5=best)", "Greenhouse gases (GHG) net emissions/removals by Land-Use Change and Forestry(LUCF) (Metric ton of carbon dioxide equivalent)", "Droughts, floods, extreme temperatures (percentage of population, average 1990-2009)" has only one value and cannot be used for future predictions using any machine learning algorithms. We have visualized various parameters using the seaborn library from fig 1. to fig 6 [4]. Table 1 contains the name of parameters and its corresponding number; in the next tables we will refer the parameter using their corresponding number.

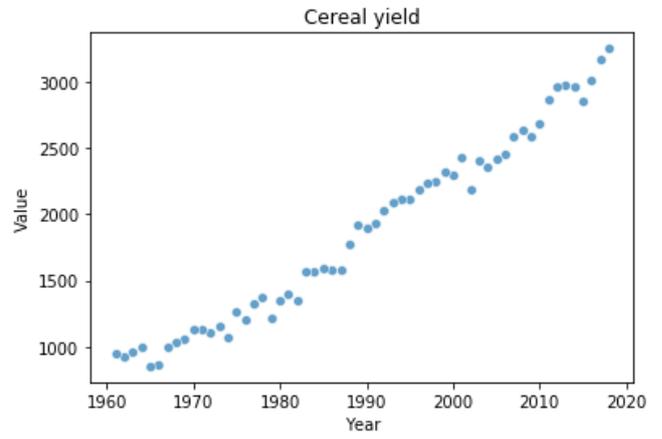

Fig. 1: Cereal yield (kilogram per hectare) per year

| Parameter number | Parameter Name |
|---|---|
| 1 | Forest area (square kilometer) |
| 2 | Agricultural irrigated land (percentage of total agricultural land) |
| 3 | Cereal yield (kilogram per hectare) |
| 4 | Access to electricity (percentage of population) |
| 5 | Carbon dioxide intensity (kilogram per kilogram of oil equivalent energy use) |
| 6 | Carbon dioxide emissions from gaseous fuel consumption (kiloton) |
| 7 | Carbon dioxide emissions (kiloton) |
| 8 | Carbon dioxide emissions from liquid fuel consumption (kiloton) |
| 9 | Carbon dioxide emissions (metric tons per capita) |
| 10 | Carbon dioxide emissions from solid fuel consumption (kiloton) |
| 11 | Total greenhouse gas emissions (kiloton of carbon dioxide equivalent) |
| 12 | Hydrofluorocarbon gas emissions (thousand metric tons of carbon dioxide equivalent) |
| 13 | Methane emissions (kiloton of carbon dioxide equivalent) |
| 14 | Nitrous oxide emissions (thousand metric tons of carbon dioxide equivalent) |
| 15 | Annual freshwater withdrawals, total (billion cubic meters) |
| 16 | Population, total |
| 17 | Urban population |

TABLE I: Parameter name and their corresponding number

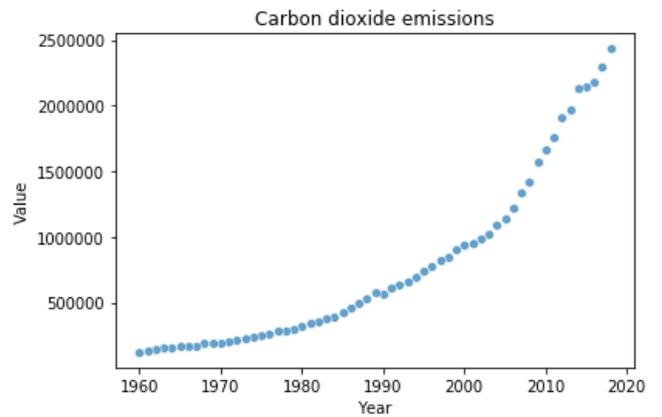

Fig. 2: Carbon dioxide emissions (kiloton) per year

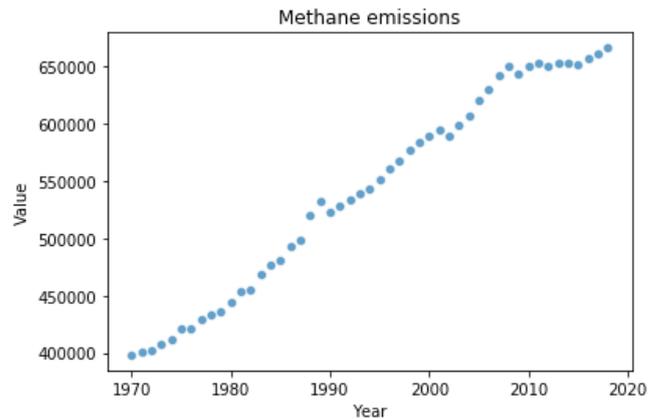

Fig. 3: Methane emissions (kiloton of carbon dioxide equivalent) per year

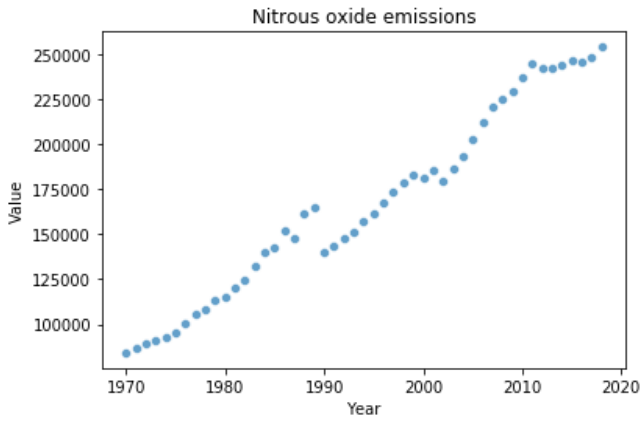

Fig. 4: Nitrous oxide emissions (thousand metric tons of CO2 equivalent) per year

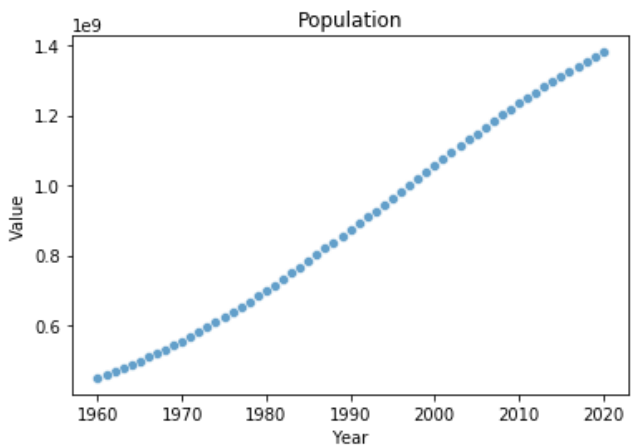

Fig. 5: Total population per year

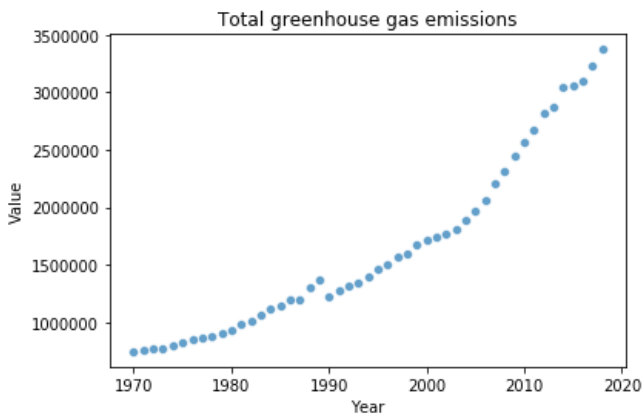

Fig. 6: Total greenhouse emissions (kiloton of carbon dioxide equivalent) per year

### B. Preprocessing

We have used Pandas library for pre-processing [5]. After dropping unwanted feature and selecting 17 important climate change parameters the values for each parameter were normalized. Normalization is a scaling technique in which values are shifted and rescaled so that they end up ranging between 0 and 1. It is also known as Min-Max scaling. Here's the formula for normalization:

$$X' = \frac{X - X_{min}}{X_{max} - X_{min}} \quad (1)$$

Here, $X_{max}$ and $X_{min}$ are the maximum and the minimum values of the feature respectively. When the value of X is the minimum value in the column, the numerator will be 0, and hence X' is 0. On the other hand, when the value of X is the maximum value in the column, the numerator is equal to the denominator and thus the value of X' is 1. If the value of X is between the minimum and the maximum value, then the value of X' is between 0 and 1

After pre-processing the data was passed through various regression algorithms – linear, polynomial, and exponential and the results were evaluated

### C. Algorithms

We have mainly used three regression algorithms, linear, polynomial and exponential using the sklearn library in python [6]. Linear regression is used for finding linear relationship between target and one or more predictors. There are two types of linear regression- Simple and Multiple. When we use one predictor to find the linear relationship between predictors and targets, we call it Simple linear Regression. When we use multiple predictors to find the linear relationship between predictors and targets, we call it Multiple Regression. We have used simple linear regression.

Polynomial Regression tries to find a nth degree curve which best fits the inputs and the targets. We have used the polynomial regression algorithm to fit a quadratic curve. Exponential Regression is used to find an exponential curve that best fits a given set of data points, below is the hypothesis function for linear regression

$$h(w_0, w_1) = w_0 + w_1 * X \quad (2)$$

Below is the hypothesis function for quadratic equation

$$h(w_0, w_1, w_2) = w_0 + w_1 * X + w_2 * X^2 \quad (3)$$

Below is the hypothesis function for exponential function

$$h(A, B) = A * B^X \quad (4)$$

where

X: The input value

h: The hypothesis function

$w_0$, $w_1$, $w_2$, A, B are trainable weights, and they are trained using gradient descent

All the trainable parameters are trained using the gradient descent algorithm. We use the hypothesis function to calculate the predicted value given the input values. The predicted values and the target values are used to calculate the loss function and we minimize the loss function using the gradient descent algorithm. Gradient descent is a first-order iterative optimization algorithm for finding a local minimum of a differentiable function.

In our case we have to minimize the loss function given below

$$J(w_0, w_1, \ldots w_n) = \frac{1}{2N} \sum_{i=1}^{N} \left( \hat{y}^{(i)} - y^{(i)} \right)^2 \quad (5)$$

where

N: Total number of samples

$\hat{y}^{(i)}$: $i^{th}$ predicted value

$y^{(i)}$: $i^{th}$ target value

### Algorithm: Gradient Descent

**Input:** Number of iterations N; n number of weights to be trained $w_0, w_1, \ldots w_n$ ; Loss function $J(w_0, w_1, \ldots w_n)$; Learning rate α

**Output:** Optimal weights which minimize the value of loss function $J(w_0, w_1, \ldots w_n)$

1. **for** i = 1 to N **do**
2. $w_n := w_n - \alpha * \frac{\partial}{\partial w_n} J(w_0, w_1, \ldots w_n)$
3. **end for**

After using gradient descent, we get optimal weights which give us minimum loss. Fig 7. to Fig 13. plot the predicted values using the hypothesis function and the actual values. The selection of one among the three models (linear, quadratic and exponential) is clearly explained in section 3 of the paper

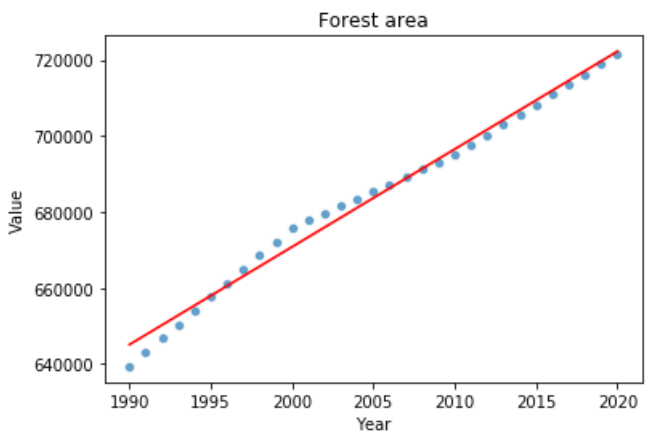

Fig. 7: Linear model for forest area

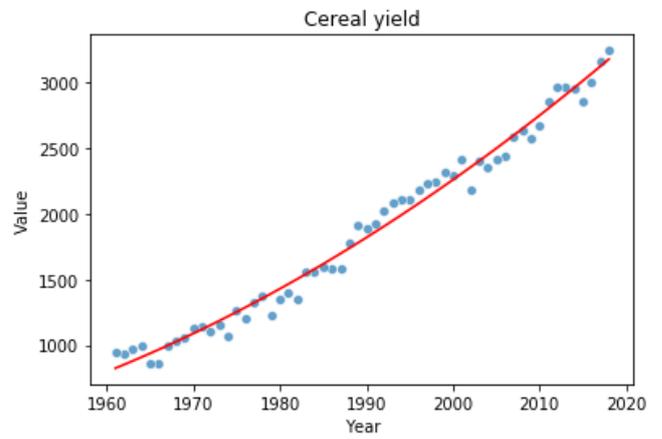

Fig. 8: Polynomial model for cereal yield

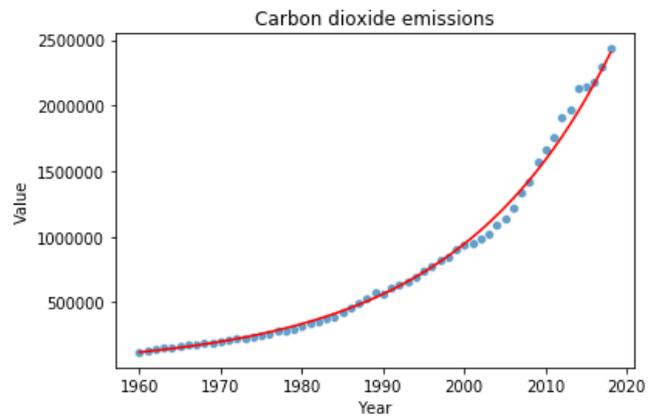

Fig. 9: Exponential model for carbon dioxide emissions

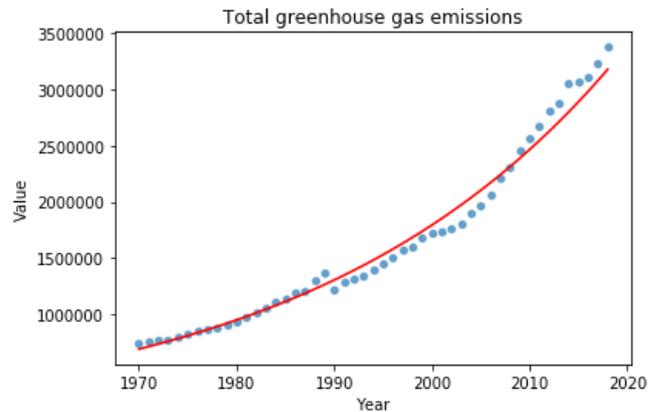

Fig. 10: Exponential model for total greenhouse emissions

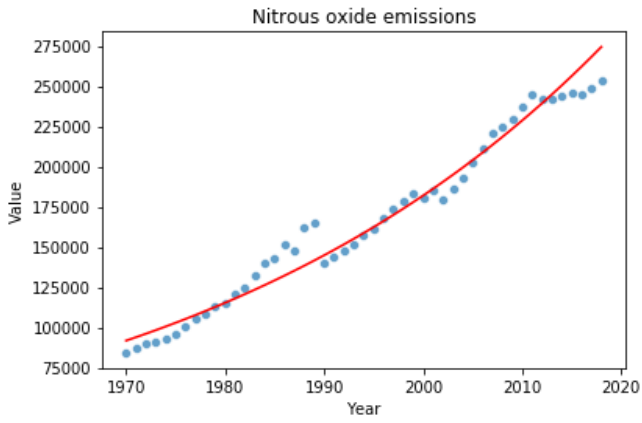

Fig. 11: Exponential model for nitrous oxide emissions

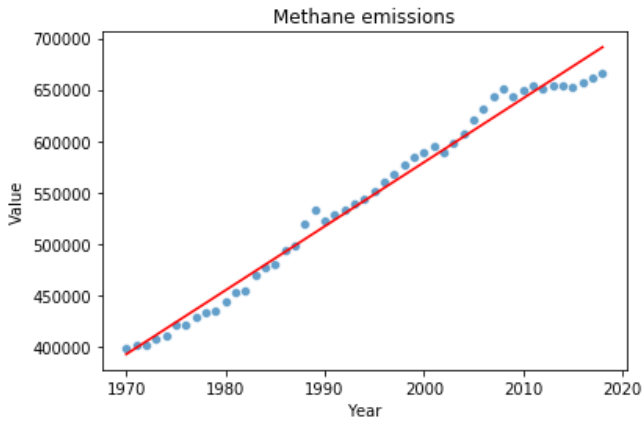

Fig. 12: Linear model for methane emissions

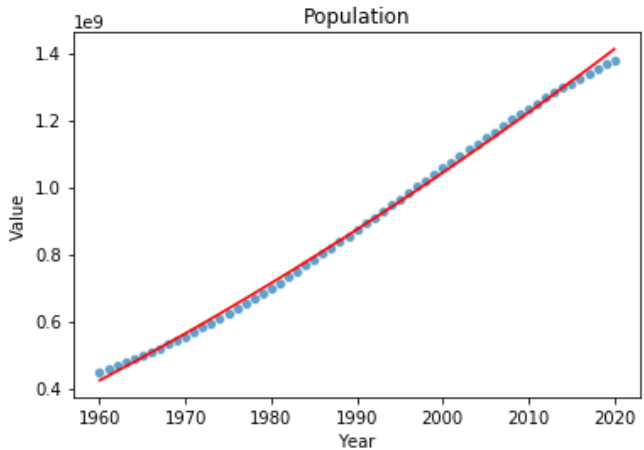

Fig. 13: Polynomial model for population

## III. EVALUATION

After fitting all three algorithms (linear, polynomial, and exponential regression) on the data we evaluate the models using two parameters, R-squared score and root mean square error. R-Squared ($R^2$ or the coefficient of determination) is a statistical measure in a regression model that determines the proportion of variance in the dependent variable that can be explained by the independent variable. In other words, r-squared shows how well the data fit the regression model (the goodness of fit). The formula of r-squared score is as follows:

$$R^2 = 1 - \frac{RSS}{TSS} = 1 - \frac{\sum_{i=1}^{n}(y_i - \hat{y}_i)^2}{\sum_{i=1}^{n}(y_i - \overline{y})^2} \quad (6)$$

where

RSS: Sum of squares due to regression (explained sum of squares)

TSS: Total sum of squares

The root-mean-square error (RMSE) is a frequently used measure of the differences between values predicted by a model and the values observed. The formula of root-mean-square error is as follows:

$$RMSE = \sqrt{\frac{\sum_{i=1}^{N}(\hat{y}_i - y_i)^2}{N}} \quad (7)$$

where

N: Total number of samples

We select the model among linear, polynomial and exponential which gives the best r-squared score and least root mean square error.

The R-squared score and root-mean-square error of the models are shown in table 2.

| Parameter number | R-squared score | Root-mean-square error | Model Type |
|---|---|---|---|
| 1 | 0.988 | 2532.218 | Linear |
| 2 | 0.912 | 0.657 | Linear |
| 3 | 0.984 | 87.071 | Polynomial |
| 4 | 0.973 | 2.298 | Polynomial |
| 5 | 0.986 | 0.0398 | Linear |
| 6 | 0.963 | 6159.992 | Polynomial |
| 7 | 0.997 | 48049.006 | Exponential |
| 8 | 0.986 | 28763.495 | Exponential |
| 9 | 0.986 | 0.065 | Exponential |
| 10 | 0.990 | 45896.375 | Exponential |
| 11 | 0.988 | 97892.936 | Exponential |
| 12 | 0.978 | 641.739 | Linear |
| 13 | 0.974 | 15899.680 | Linear |
| 14 | 0.975 | 8187.562 | Exponential |
| 15 | 0.861 | 42.361 | Exponential |
| 16 | 0.998 | 12983761.946 | Polynomial |
| 17 | 0.999 | 1578542.742 | Polynomial |

TABLE II: R-squared score and root-mean-square error of the models

## IV. INFERENCE

The trained models are used to predict the values of the 17 parameters for the year 2025, 2030 and 2035. For linear model we only pass the year value for example 2025 to predict its value using the linear model. For polynomial model we pass two values, the year value and its square, 2025 and $2025^2$. For the exponential model we pass the year value for example 2025 and get the natural logarithm of the predicted value. The predicted values are shown in table 3.

| Parameter number | 2025 | 2030 | 2035 |
|---|---|---|---|
| 1 | 735124.032 | 747984.959 | 760845.887 |
| 2 | 42.797 | 45.254 | 53.090 |
| 3 | 3580.979 | 3882.175 | 4195.809 |
| 4 | 115.022 | 132.034 | 150.925 |
| 5 | 2.813 | 2.949 | 3.085 |
| 6 | 147058.578 | 174585.037 | 204456.594 |
| 7 | 3454882.1580 | 4477007.864 | 5801529.111 |
| 8 | 1099238.781 | 1467701.883 | 1959673.235 |
| 9 | 2.051 | 2.408 | 2.826 |
| 10 | 2282348.274 | 2932672.651 | 3768298.195 |
| 11 | 3965426.377 | 4647366.025 | 5446579.740 |
| 12 | 22812.131 | 25822.286 | 28832.440 |
| 13 | 769936.739 | 816603.168 | 866098.084 |
| 14 | 279436.016 | 297558.130 | 315680.245 |
| 15 | 800.522 | 853.403 | 909.777 |
| 16 | 1.509e+09 | 1.610e+09 | 1.713e+09 |
| 17 | 5.357e+08 | 5.936e+08 | 6.548e+08 |

TABLE III: Predicted values

## V. CONCLUSION

From the various graphs discussed above we have observed the upsurge of climatic changes and therefore must raise awareness and understand the seriousness of this rise. The linearly, polynomially, and exponentially increasing graphs clearly state the increasing trends of all the parameters for different years. It can thus prove to be very useful to help and save many sectors of India. Deep-learning algorithms can be used with this model to enhance the results. The idea can be further extended by publishing this data on websites to help people of India understand the seriousness of the problem.

## REFERENCES

[1] Picciariello, A., Colenbrander, S., Bazaz, A. and Roy, R. (2021) The costs of climate change in India: a review of the climate-related risks facing India, and their economic and social costs

[2] Anil Agarwal. Climate change : A challenge to India's economy, Centre for Science and Environment

[3] https://data.humdata.org/dataset/world-bank-climate-change-indicators-for-india

[4] Waskom, M., Botvinnik, Olga, O'Kane, Drew, Hobson, Paul, Lukauskas, Saulius, Gemperline, David C, ... Qalieh, Adel. (2017). mwaskom/seaborn: v0.8.1 (September 2017). Zenodo. https://doi.org/10.5281/zenodo.883859

[5] McKinney, W., & others. (2010). Data structures for statistical computing in python. In Proceedings of the 9th Python in Science Conference (Vol. 445, pp. 51–56).

[6] Pedregosa, F., Varoquaux, Ga"el, Gramfort, A., Michel, V., Thirion, B., Grisel, O., ... others. (2011). Scikit-learn: Machine learning in Python. Journal of Machine Learning Research, 12(Oct), 2825–2830.